\documentclass{article}
\usepackage{arxiv}

\usepackage[utf8]{inputenc} % allow utf-8 input
\usepackage[T1]{fontenc}    % use 8-bit T1 fonts
\usepackage{hyperref}       % hyperlinks
\usepackage{url}            % simple URL typesetting
\usepackage{booktabs}       % professional-quality tables
\usepackage{amsfonts}       % blackboard math symbols
\usepackage{nicefrac}       % compact symbols for 1/2, etc.
\usepackage{microtype}      % microtypography
\usepackage{lipsum}

\usepackage{graphicx}       % This should be added to add images   
\usepackage{subcaption}
\usepackage[table,xcdraw]{xcolor}

\usepackage{comment}

\title{DeepFilter: an ECG baseline wander removal filter using deep learning techniques}

\author{
  Francisco P. Romero \\
  Department of Electrical Engineering\\
  Polytechnique Montreal\\
  \texttt{fperdigon88@gmail.com} \\
  \thanks{Thanks to Liset Vazquez Romaguera for fruitful discussions and document review.}
   \And
  David C. Piñol \\
  Department of Electrical, Electronics \\
  and Communication Engineering\\ 
  Friedrich-Alexander-Universität\\
  Erlangen-Nürnberg (FAU)\\
  \texttt{david.pinol@fau.de} \\
   \AND
  Carlos R. Vázquez-Seisdedos\\
  Center of Neuroscience and Signal Processing\\
  Universidad de Oriente\\
  cvazquez@uo.edu.cu\\
}

\begin{document}
\maketitle

\begin{abstract}
According to the World Health Organization, around 36\% of the annual deaths are associated with cardiovascular diseases and 90\% of heart attacks are preventable. Electrocardiogram signal analysis in ambulatory electrocardiography, during an exercise stress test, and in resting conditions allows cardiovascular disease diagnosis. However, during the acquisition, there is a variety of noises that may damage the signal quality thereby compromising their diagnostic potential. The baseline wander is one of the most undesirable noises. In this work, we propose a novel algorithm for BLW noise filtering using deep learning techniques. The model performance was validated using the QT Database and the MIT-BIH Noise Stress Test Database from Physionet. In addition, several comparative experiments were performed against state-of-the-art methods using traditional filtering procedures as well as deep learning techniques. The proposed approach yields the best results on four similarity metrics: the sum of squared distance, maximum absolute square, percentage of root distance, and cosine similarity with $4.29 \pm 6.35$ au, $0.34 \pm 0.25$ au, $45.35 \pm 29.69$ au and, $ 91.46 \pm 8.61$ au, respectively. The source code of this work, containing our method and related implementations, is freely available on Github.
\end{abstract}

% keywords can be removed

\section{Introduction}
For many years, cardiovascular diseases (CVDs) has remained as the main leading cause of sudden cardiac death worldwide. The World Health Organization (WHO) estimate that 17.9 million deaths every year are attributed to CVDs, which represents 31\% of the global deaths \cite{CVDs}. Heart attack and strokes constitute 85\% of these deaths.
Nonetheless, recent studies have shown that 90\% of heart attacks could be preventable \cite{kullo2016incorporating}. Early and effective diagnosis is essential to prevent cardiac attacks. The electrocardiogram (ECG) is a well-established  technique for the diagnostic of heart diseases. It consists of acquiring the electrical activity of the heart captured over time by an external electrode attached to the skin.The ECG can estimate the physical heart condition and detect a wide variety of abnormalities such as arrhythmias, coronary artery block, among others. t can be acquired under different conditions: resting, ambulatory electrocardiography, and during an exercise stress test  \cite{Non-Invasive_AHA}. The last two allow a better diagnosis but are strongly affected by different noise sources namely: 60/50 Hz power line noise, contact noise, patient-electrode motion artifacts, EMG noise and baseline drift \cite{moody1984noise}. 
The ambulatory electrocardiography (also known as Holter Monitoring or Ambulatory ECG) is presented when the electrical activity of the heart is recorded during daily activities. It usually takes 24 hours to record or more. The likelihood of encountering heart events is higher than the rest recording conditions. 
The exercise stress test (also known as Treadmill Test, Exercise Test, Exercise Cardiac Stress Test, and ECST) records the heart function (usually between 5 - 15 minutes) while the subject walks in place, performs an effort in a bicycle or on an electric treadmill. Many aspects of the heart function can be checked including heart rate, breathing, blood pressure, and how tired the patient becomes when exercising. Despite the increased probability of a better heart event detection in ECG long-term recording, new challenges come across. For instance, the ambulatory or stress test ECG are prone to be highly contaminated with noise given their acquisition conditions \cite{sornmo2005bioelectrical}. Amongst the variety of noises, the baseline wander (BLW) is one of the most challenging and is a serious limitation during stress tests and ambulatory ECG. 

The baseline wander, also known as baseline drift, is a low-frequency noise constituted by frequencies components ranging between 0.05 Hz and 3 Hz \cite{sornmo2005bioelectrical}. The time base axis of the ECG signal seems to ‘wander’ and down instead of be straight. The visual effect consists in a movement of the whole signal from its normal base. The BLW is caused by misplaced electrodes (electrode-skin impedance), bad skin preparation, patient’s movement and breathing, scarce contact between electrode cables, and a combination of the aforementioned factors. It negatively affects the heart events detection in automatic diagnosis algorithms, yielding a useless outcome in the majority of cases. Consequently, removing this noise is crucial to guarantee a proper clinical assessment using ECG recording. 

Several solutions have been proposed to mitigate the BLW. Different authors have proposed a wide variety of techniques such as classical digital filters \cite{kumar2015removal}, Wavelet transforms \cite{tinati2005ecg}, empirical mode decomposition \cite{blanco2008ecg}, among others. Romero et al. \cite{romero2018baseline} presented a comparative study between many of these approaches. While some methods obtained acceptable results, the morphological deformation in the ECG signal is still an outstanding problem.
Despite BLW noise filtering being a research topic widely studied as previously stated, it is still an open problem that we consider is worth it to revisit with a renewed vision. This will allow seeking better BLW filters by looking at state-of-the-art techniques such as artificial intelligence driven filters.

In this paper is proposed a novel algorithm that filters the baseline wander noise using deep learning techniques. It was used the QT Database and the MIT-BIH Noise Stress Test Database from Physionet and several experiments were performed, an ablation of different models was made in order to determine the importance of linear non-linear activations, and the of advantages of using dilated convolutions.
The rationale of using deep learning for BLW removal is that we hypothesize that it is possible to learn “smart filters” that know what is the desired output (the ECG signal) and the undesired result (noise). Several deep filters will learn how to properly filter small sections of the input signal while conserving the ECG signal morphology by the use of a similarity loss function.
This paper is composed of several sections. Section II provides information regarding the databases used. In section III is written the pre-processing techniques and the deep learning models developed. Results and discussion can be depicted in section IV and conclusion in section V.
In order to maximize reproducibility the source code of this work will be freely available in GitHub [\href{https://github.com/fperdigon/DeepFilter}{https://github.com/fperdigon/DeepFilter}]. Also for educational purposes this work can be run on Google Colab where free GPU interactive sessions using jupyter notebooks can be accessed 
[\href{https://colab.research.google.com/drive/1S1HjkQnrA0EbEDJFr0D-6DAt62RCcP5_?usp=sharing}{Colab link}].

\section{Related works}
BLW noise removal has a wide range of literature research behind and different techniques have emerged as alternatives to this problem. The works presented in \cite{sonali2013patial} and \cite{romero2018baseline} gather a collection of prevalent BLW noise removal techniques. State-of-the-art techniques cover methods such as Adaptive filtering, Wavelet adaptive filtering, Moving average, Finite and Infinite impulse filter response, cubic spline, Independent components analysis, among others. Most of these traditional methods have shown good performance. However, the comparative study in \cite{romero2018baseline} showed that Finite Impulse Response (FIR) and Infinitive Impulse Response (IIR) filters poses the best results in BLW noise mitigation according to different signal distortion metrics.
Furthermore, as long as deep learning techniques take place in a variety of signal processing applications, it begins to appear remarkable works inside ECG denoising applications. The work presented in \cite{antczak2018deep} uses a transfer learning technique by pre-training the network with synthetic data and fine-tuning with real data. In addition, it is also notorious the work of \cite{chiang2019noise}, in which their proposal consists of a denoising autoencoder (DAE) based on a fully convolutional network (FCN) which yields interesting results to be taken into account. Similarly, another approach that also uses a DAE to enhance the ECG signal is presented in \cite{xiong2015denoising}.

\section{Materials}
All the experiments performed in this paper used the QT Database \cite{laguna1997database} from Physionet \cite{goldberger2000physiobank}. This dataset contains real ECG records that represent a wide range of QRS and ST-T morphologies in all the possible variability. It is composed of 105 ECG signals with 15-min and two channels each one sampled at 250 Hz. The experiments performed in this paper used all the records available in this dataset. 
In order to add real baseline drifts to the QT database, real BLW signals from the MIT-BIH Noise Stress Test Database (NSTDB)\cite{moody1984noise} were used. The database contains 30 minutes of 12 ECG recordings and 3 recordings of typical noise in stress tests at 360 Hz sampling frequency. These noises are baseline wander produced by the patient's breathing, muscle artifact and electrode motion artifact. The ECG records are randomly corrupted with the noise present in the three noise channels. The noise records were recorded during physical stress tests, with electrodes placed on the limbs in positions where ECG cannot be acquired. In this paper, it was only used records with BLW noise generated by respiration and electrode motion artifacts.

\section{Methods}
In this section we present our deep learning BLW filtering solution (DeepFilter). The proposed model is a fully convolutional architecture based on multipath modules where the input is the ECG contaminated with BLW and the output is the ECG without noise. Fully convolutional networks (FCN) are widely used for tasks where the output dimensions have to be the same as the input dimension. As application examples of FCN we can mention U-net for semantic segmentation \cite{ronneberger2015u}, VoxelMorph for deformable medical image registration \cite{balakrishnan2019voxelmorph}, and black and white image colorization \cite{zhang2016colorful} for images restoration to mention some.

\begin{figure}[!h]
    \centering
    %\fbox{\rule[-.5cm]{0cm}{4cm} \rule[-.5cm]{4cm}{0cm}}
    \includegraphics[scale=0.65]{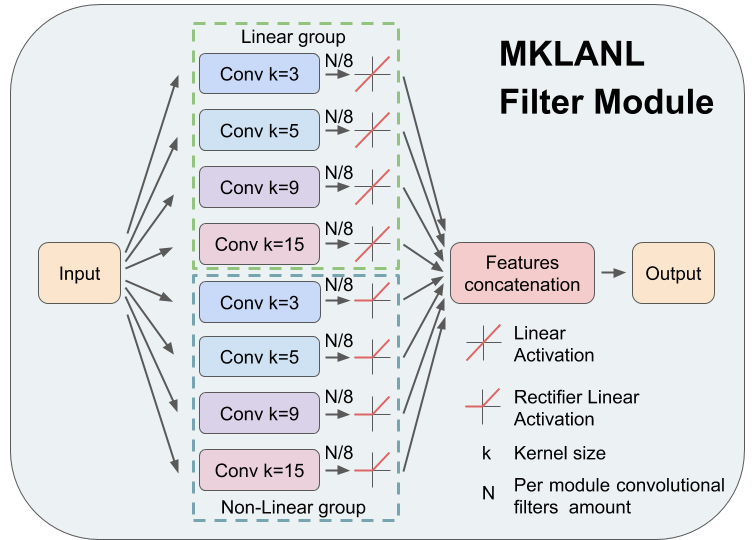}
    \caption{Proposed Multi Kernel Linear And Non-Linear Filter Module.}
    \label{fig:deep_filter_layer}
\end{figure}

Our approach uses multipath modules, which place different convolutional layers at the same level and let the backpropagation algorithm choose not only the weights but also the best path for the signal to pass through. Figure \ref{fig:deep_filter_layer} shows the proposed Multi Kernel Linear And Non-Linear Filter Module (MKLANL Filter Module) which is inspired by the Inception module first introduced by Szagedy and collaborators in  \cite{szegedy2015going}. In deep learning models one of the hyperparameters that are hard to set is kernel size. To find the best value at each layer it is common to perform several experiments. A better way is to use multipath modules like Inception (or ours). Then for each level the best kernel will be learned.

The proposed MKLANL Filter Module is composed by two internal groups, the Linear group, and the Non-Linear group. Each internal group contains four types of convolutional layers with kernels equal to (3, 5, 9, and 15) followed whether by a linear activation or rectified linear unit (ReLU) , depending on the group. The number of convolutional filters for each type is N/8, where N is the hyperparameter that controls the total amount of filters per multipath module. Generally, N is referred as the amount of extracted features since the result of each convolution operation has associated a transformed output signal. The rationale behind having convolution with linear and non-linear activations is the same as having different convolutional kernels: let the model choose during training which path is better and how much each of them will contribute to the output.

\begin{figure}[htb]
    \centering
    %\fbox{\rule[-.5cm]{0cm}{4cm} \rule[-.5cm]{4cm}{0cm}}
    \includegraphics[scale=0.78]{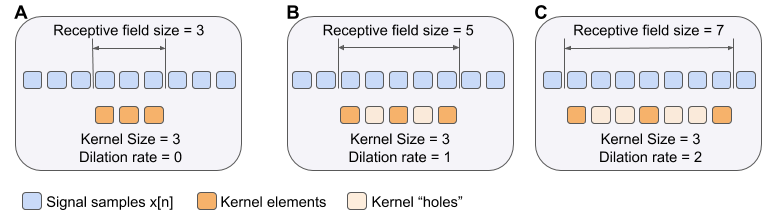}
    \caption{Dilated convolution.  \textbf{A} is a dilated convolution with dilation rate r = 0, equivalent to the standard convolution. \textbf{B} has dilation rate r = 1 and \textbf{C} a dilation rate r = 2. All dilated convolution have a, equal kernel size of 3, the same number of parameters, and the same computational cost.
}
    \label{fig:dilated_conv}
\end{figure}

The effect of low-frequency signals is better appreciated when a large amount of consecutive samples is under analysis. That is the main reason why good low frequency filters have large windows. The analog procedure in convolutional networks will be increasing the kernel size, however, this also increments the computational load. Dilated convolutions are a smart and elegant solution to this problem, by using non-consecutive kernels \cite{yu2015multi}. This solution allows to increase the kernels receptive field while the computational load remains the same (see Figure \ref{fig:dilated_conv}).

For a one dimensional signal \(x[i]\), the output \( y[i]\) of a dilated convolution operation with a dilation rate r and a filter mask \(w[s]\) is defined by the following formula:

\begin{equation}
y\left [ i \right ]=\sum_{s=1}^{S}x\left [ i + r \cdot s \right ]w\left [ s \right ]
\label{eq:d_conv}
\end{equation}

\begin{figure}[h]
    \centering
    %\fbox{\rule[-.5cm]{0cm}{4cm} \rule[-.5cm]{4cm}{0cm}}
    \includegraphics[scale=0.65]{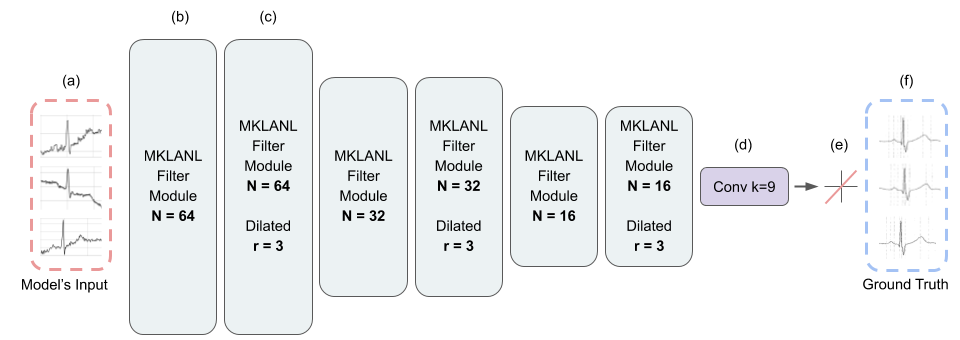}
    \caption{DeepFilter model.}
    \label{fig:prop_arch}
\end{figure}

Figure \ref{fig:prop_arch} illustrates the proposed deep learning architecture. It is composed by six MKLANL Filter Modules arranged in a sequence. The first two MKLANL Filter Modules (b) and (c) have a total of 64 convolutional filters internally arranged as shown in Figure \ref{fig:deep_filter_layer}. In the second multipath module (c) the convolutional operations were set with a dilation rate r = 3 enabling dilated convolution. This same dilation rate is also set on the fourth and sixth modules. The number of extracted features decreases along the network starting with 64 on modules one and two, then 32 features on modules three and four, ending with 16 features for modules five and six. The final step is one convolutional filter with kernel = 9  which conforms the output signal. Since ECG are bipolar signals, linear activation was used for this final step thereby allowing the output to have either positive or negative values.

The model’s input signals (See box (a) \ref{fig:prop_arch}) are ECG signals contaminated with the BLW noise from the NSTDB. The used ground truths are the same ECG signals used as input but without BLW noise contamination. In this way the model will learn through the iterative back propagation process the filters kernels values for BLW removal with a low distortion.

For the optimization process we used as a surrogate loss function (\(L_{filtering}\)) a combination of two metrics: the Sum of Squared Distance (SSD) and the Maximum Absolute Distance (MAD). We consider them to represent better the task we want to perform. The SSD term (Equation \ref{eq:loss} left term) describes the signals difference along sample dimension and does not minimize the effect of outlier values as averaged metrics such as Mean Squared Error (MSE) or the Signal to Noise Ratio (SNR). The MAD term (Equation \ref{eq:loss} right term) calculates the signals difference maximum value. This term enforces the model to minimize the difference with the original signal point wise while the SSD term analyses the whole signal. Since both losses have a different values range we used a balance term (\(\lambda\)). We empirically found \(\lambda\)=50 to be a value that works well in our current deep learning setup.

\begin{equation}
L_{filtering}=\sum_{m=1}^{r}(y_{true}(m)-y_{pred}(m))^{2} + \lambda \times max\left | (y_{true}(m)-y_{pred}(m))^{2} \right |_{m=1}^{r}
\label{eq:loss}
\end{equation}

where:\\
\(y_{true}\): Original signal (ECG without noise) \\
\(y_{pred}\): Model predicted signal (ECG filtered by the model)\\
\(\lambda\): Balance term \\
\(m\): Sample variable \\
\(r\): Maximum signal size \\

\section{Experiments and Results}

\subsection{Experiments design}

\begin{figure}[htb]
    \centering
    %\fbox{\rule[-.5cm]{0cm}{4cm} \rule[-.5cm]{4cm}{0cm}}
    \includegraphics[scale=0.77]{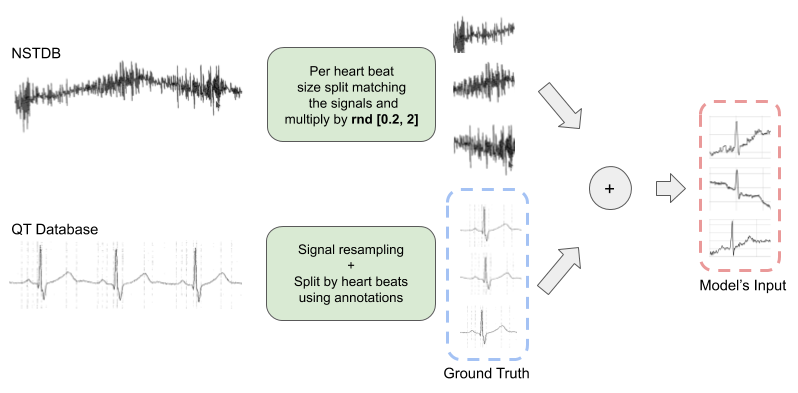}
    \caption{Data preprocessing pipeline.}
    \label{fig:data_pip}
\end{figure}

For our experiments we designed an BLW removal benchmark where different methods can be evaluated using the same test set under the same conditions. See \ref{fig:data_pip} for the data preprocessing pipeline.

For all the 105 signals on the QT Database were over-sampled from 250 Hz to 360 Hz in order to match the NSTDB sampling frequency. Heartbeats were extracted using the specialist’s annotations. We detected a small number of wrong annotations regarding the beat start/end. This causes two consecutive beats to be considered as one. To mitigate this error we discard beats bigger than 512 samples (1422.22 ms). 

We separated the heartbeats belonging to 14 signals (see Table \ref{tab:test_set}) to be used as a test set representing the ~13\% percent of the total signals amount. Two signals from each of the seven datasets used to create the QT Database. This will ensure we have different pathologies on the test set giving a better idea of the generalization capability of the methods under evaluation.

\begin{table}[!h]
\caption{Signals used for test set representing all internal data on the QT dataset.}
\label{tab:test_set}
\setlength{\tabcolsep}{6pt} % Default value: 6pt
\renewcommand{\arraystretch}{1.25} % Default value: 1
\resizebox{\textwidth}{!}{%
\begin{tabular}{ccccccc}
\rowcolor[HTML]{EFEFEF} 
\textbf{\begin{tabular}[c]{@{}c@{}}MIT-BIH\\ Arrhythmia\\ Database\end{tabular}} & \textbf{\begin{tabular}[c]{@{}c@{}}MIT-BIH ST\\ Change\\ Database\end{tabular}} & \textbf{\begin{tabular}[c]{@{}c@{}}MIT-BIH\\ Supraventricular\\ Arrhythmia\\ Database\end{tabular}} & \textbf{\begin{tabular}[c]{@{}c@{}}MIT-BIH\\ Normal\\ Sinus Rhythm \\ Database\end{tabular}} & \textbf{\begin{tabular}[c]{@{}c@{}}European\\ ST-T\\ Database\end{tabular}} & \textbf{\begin{tabular}[c]{@{}c@{}}‘Sudden death’\\ patients from\\ BIH\end{tabular}} & \textbf{\begin{tabular}[c]{@{}c@{}}MIT-BIH\\ Long-Term\\ ECG\\ Database\end{tabular}} \\
sel123 & sel302 & sel820 & sel16420 & sele0106 & sel32 & sel14046 \\
sel233 & sel307 & sel853 & sel16795 & sele0121 & sel49 & sel15814
\end{tabular}
}
\end{table}

Noise from NSTDB containing BLW due to breathing and electrode movement (signal named \textit{em} on the database) was used to contaminate the ECG signals. The \textit{em} record was split to match the beat samples length. Channels 1 and 2 were concatenated after saving the 13\% of each channel signal length to contaminate the signals previously designated for the test set.

Having a separate signal test set will ensure the reliability of the results when using learning algorithms. This also will avoid learning algorithms having any sort of advantage when comparing with other classical methods.

For injecting the noise we followed the same approach used by the NSTDB in \cite{moody1984noise} where the noise is randomly injected with values from 0.2 up to 2 times the ECG signal maximum peak value.

\subsection{Implementation of comparative methods}
\label{section:comparative_methods}

We implemented state-of-the-art approaches to compare their performance with the proposed method. As representation of non-learnable approaches a FIR and IIR filters were used in our experiments. These classical filters obtained best results on a comparative study performed by Romero and collaborators in \cite{romero2018baseline}. Also, we implemented two deep learning state of the art methods for ECG denoising. The first one based on Deep Recurrent Neural Networks (DRNN) \cite{antczak2018deep} and the second one based on Fully Convolutional Denoising Auto-Encoder (FCN-DAE)  \cite{chiang2019noise}. Finally, with the objective to show the relevance of non-linear activations in deep learning models, the advantages of using multi branch architectures and the importance of dilated convolutions, we implemented three models for an ablation study.

\textbf{FIR filter:} One of the methods that has reported high accuracy in removing the BWL is a classical high-pass FIR digital filter presented in \cite{van1985removal}. It has been depicted in a comparative study as a strong candidate in removing BWL (Romero et al., 2019). The authors calculated this filter with 28 length Kaiser window and 51 coefficients. Nevertheless, in this work, this filter was redesigned taking into account more computational resources and these are restrictions we used for our implementation: the width transition from stop band to pass band 0.07 and an attenuation in the stop band of 30.5 dB. As a result, the design parameters were an order of 8079 and a beta of 2.18. In addition, it is important to highlight that the chosen cutoff frequency was 0.67Hz, which has been proven a proper high-pass frequency for BWL removal according to the AHA \cite{klingfield2007recommendations} . 

\textbf{IIR filter:} Another remarkable classical filter that has stood out in state of the art literature and experiments (Romero et al., 2019) is an IIR high pass filter proposed in (Kumar et al., 2015). The main advantage of this IIR filter is that it requires less computational resources and at the same time it is able to attain high accuracy as its FIR filter counterpart. The design is composed by an order of 4 and a Butterworth type filter. Moreover, in order to avoid the common phase delay in the signal output, it is crucial to notice that, bi-directional filters are used in both filters \cite{oppenheim1999discrete}. These filters achieve a zero-phase distortion by evaluating the input signal in both the forward and reverse directions.

\textbf{Deep recurrent neural network (DRNN):} The method based on recurrent networks proposed by Antczak \cite{antczak2018deep} was implemented. As described on the paper we used a LSTM module with 64 units, the \textit{return\_sequence} parameter was set to \textit{True} since this allows to return the full sequence (this means that the output tensor will have the same shape as the input tensor). The LSTM is followed by two fully connected layers (also called Dense layers) of 64 neurons each, the used activation was relu. The final layer is composed of a Dense layer with one neuron using a linear activation. The used loss function is the MSE as the original paper \cite{antczak2018deep}.

\textbf{FCN-DAE:} Denoising autoencoders allows learning architectures to automatically extract the most relevant features for the further reconstruction in presence of noise at the input by exploiting the generative capabilities of auto encoder models  \cite{vincent2008extracting}. Chiang and collaborators \cite{chiang2019noise} used a fully convolutional denoising auto encoder, a convolutional based architecture that follows the same principles as original autoencoders. The authors included batch normalization for better gradient flow and used the MSE as a loss function.

\textbf{Ablation study:} To assess the influence of different architecture variations we implemented three different models. Al models share the same number of convolutional filters at the same level of providing a similar architectural root. All models were tested using the proposed loss and the training protocol which will be explained in the next section. Table \ref{tab:model_ablation} shows the used architecture for the aforementioned models. The \textit{Vanilla FCN Linear} model used linear activation after all the convolutional filters. However the \textit{Vanilla FCN Non-Linear} uses ReLU a non-linear activation in all its internal filters but the last one. Finally, the \textit{Multibranch model} uses the MKLANL filter module introduced in this work but dilation rate is 0 (no dilation).

\begin{table}[!h]
\centering
\caption{Models for the ablation study}
\label{tab:model_ablation}
\setlength{\tabcolsep}{6pt} % Default value: 6pt
\renewcommand{\arraystretch}{1.25} % Default value: 1
\begin{tabular}{c c c}
%\begin{tabular}{|c|c|c|}
\rowcolor[HTML]{EFEFEF} 
\textbf{Vanilla FCN Linear} & \textbf{Vanilla FCN Non-Linear} & \textbf{Multibranch} \\
64 Conv, k=9, linear & 64 Conv, k=9, ReLU & 64 MKLANL Filter Module \\
64 Conv, k=9, linear & 64 Conv, k=9, ReLU & 64 MKLANL Filter Module \\
32 Conv, k=9, linear & 32 Conv, k=9, ReLU & 32 MKLANL Filter Module \\
32 Conv, k=9, linear & 32 Conv, k=9, ReLU & 32 MKLANL Filter Module \\
16 Conv, k=9, linear & 16 Conv, k=9, ReLU & 16 MKLANL Filter Module \\
16 Conv, k=9, linear & 16 Conv, k=9, ReLU & 16 MKLANL Filter Module \\
1 Conv, k=9, linear & 1 Conv, k=9, linear & 1 Conv, k=9, linear
\end{tabular}
\end{table}

\subsection{Training Protocol}

The training protocol explained in this section was applied to all deep learning based methods.

\textbf{Train/Validation/Test split:} As previously explained we separated the heartbeats belonging to 14 signals to be used as a test set (two from each dataset used to conform the QT Database) . This represents the ~13\% percent of the total signals amount. The heart beats from the remaining 91 signals were randomly shuffled and splited to be used as a train set (70\%) and validation set (30\%).

\textbf{Batch size: }A batch size of 32 signals was used for training. This brings the advantage of the speed-up of matrix-matrix products making the training faster  \cite{bengio2012practical}.
Optimization and Learning Rate Scheduling: Adam optimization algorithm \cite{kingma2014adam} was used with an initial learning rate of \(10^{-3}\). This learning rate was reduced by a factor of 2 after 2 epochs without improvements in the validation set SSD metric. The allowed minimum learning rate was \(10^{-10}\).

\textbf{Iterations:} The initial number of epochs was \(10^5\). Due to the early stopping criterion that was implemented the final number of epochs is automatically determined during training. For early stopping, the validation set SSD metric was monitored, and stopped after 10 epochs if no improvements were detected.

\subsection{Evaluation Metrics}
According to studies published in \cite{manikandan2008ecg, nygaard2001rate}, distance-based metrics are the most suitable for the evaluation of signal similarity. In order to evaluate the performance of methods in removing the BLW minimizing the distortion on the  ECG signal morphology we used: Absolute maximum distance (MAD), sum of the square of the distances (SSD), percentage root-mean-square difference (PRD),Cosine Similarity. The following paragraphs selected metrics are explined.

\textbf{Maximum absolute distance (MAD):} This metric was used in \cite{nygaard2001rate} in a compression context in ECG signals. It has been considered a well-known similarity metric to quantify the ECG quality after being processed. It measures the maximum absolute distance between the original and filtered signal. The formula is given by:

\begin{equation}
 MAD(s_{1},s_{2})= max\left | s_{1}(n)-s_{2}(n) \right | \:\:\:\:\: 1\leq n\leq N 
\end{equation}

Where \(s_{1}\) and \(s_{2}\) are the original signal and the filtered one to be compared, \(n\) is the index of the current sample and \(N\) is the length of the signals. Both signals must have the same length. The same applies to Sum of the square of the distances and Percentage root-mean-square difference formulas.

\textbf{Sum of the square of the distances (SSD):} This is another popular similarity metric \cite{nygaard2001rate}, which measures the sum of squared distances between the original and the filtered signal. It provides an idea of how similar the signals are along their entire duration. The formula is given by:

\begin{equation}
 SSD(s_{1},s_{2})=\sum_{n=1}^{N}(s_{2}(n)-s_{1}(n))^{2} \:\:\:\:\: 1\leq n\leq N 
\end{equation}

\textbf{Percentage root-mean-square difference (PRD): }This is a common similarity metric based on distance and its value is given in percentage.  The objective is to obtain a smaller value, and consequently, the filtered signal is considered the best approximation of the original. The formula is the following:

\begin{equation}
 PRD(s_{1},s_{2})=\sqrt{\frac{\sum_{n=1}^{N}(s_{2}(n)-s_{1}(n))^{2}}{\sum_{n=1}^{N}(s_{2}(n)-\overline{s_{1}})^{2}}}*100\% \:\:\:\:\: 1\leq n\leq N 
\end{equation}

\textbf{Cosine similarity:} This is a measure of similarity between two vectors. It is a normalized bounded inner product by L2 norms. The cosine similarity comes because the normalized dot product by the Euclidean L2 normalization is the cosine of the angle between the points denoted by the vectors in the unit sphere \cite{schutze2008introduction}. The formula is given by:

\begin{equation}
 CosSim(s_{1},s_{2})=\frac{\sum_{n=1}^{N}s_{1}(n)s_{2}(n)}{\sqrt{\sum_{n=1}^{N}s_{1}^{2}(n)} \sqrt{\sum_{n=1}^{N}s_{2}^{2}(n)}} =\frac{\left \langle \mathbf{s_{1}},\mathbf{s_{2}} \right \rangle}{\left \| \mathbf{s_{1}} \right \| \left \| \mathbf{s_{2}} \right \|} \:\:\:\:\: 1\leq n\leq N 
\end{equation}

Where \(s_{1}\) and \(s_{2}\) are the original signal and the filtered one to be compared, \(n\) is the index of the current sample and \(N\) is the length of the signals. It is not necessary that both signals feature the same length, however in our particular use case signals length is always the same.

It is important to notice that as closer the cosine similarity is from 1, therefore more similar are the two vectors or, in our case, the two signals.

\subsection{Results and discussion}

We now present a comparison between the proposed method, state-of-the-art and two classical approaches that yielded good results on previous baseline wander comparison studies. As explained in Section \ref{section:comparative_methods}, FIR and IIR filters were implemented inspired by Van Alste et al. \cite{van1985removal} and  Kumar et al. \cite{kumar2015removal} papers, respectively. Also we implemented two recently proposed deep learning based methods, the first using DRNN \cite{antczak2018deep}, the second based on denoising auto encoder architectures \cite{chiang2019noise}.

Our experiments are aimed at investigating three main aspects: (1) find the model that introduces less signal deformation, (2) determine the influence of linear and non-linear on the model’s final performance, and (3) determine the influence of dilated convolutions. We leverage the signal deformation using the following similarity metrics: Sum of Squared Distance (SSD), Maximum Absolute Distance, Percent of root-mean-square Distance (PRD), and Cosine Similarity. Statistical significance was calculated by applying a Wincoxon signed-rank test using the Scipy library implementation. P < 0.01 was considered to indicate statistical significant difference.

\begin{table}[!h]
\centering
\caption{Methods’s performance result}
\label{tab:test_results}
\setlength{\tabcolsep}{6pt} % Default value: 6pt
\renewcommand{\arraystretch}{1.25} % Default value: 1
\begin{tabular}{
>{\columncolor[HTML]{FFFFFF}}l cccc}
\hline
\textbf{Method/Model} & \cellcolor[HTML]{FFFFFF}\textbf{SSD (au)} & \cellcolor[HTML]{FFFFFF}\textbf{MAD (au)} & \cellcolor[HTML]{FFFFFF}\textbf{PRD (\%)} & \cellcolor[HTML]{FFFFFF}\textbf{Cosine Sim $\times100$ (\%)} \\ \hline
FIR Filter & 44.97 $\pm$85.03 & 0.69 $\pm$0.58 & 65.77 $\pm$21.81 & 69.70 $\pm$21.01 \\
IIR Filter & 35.63 $\pm$69.77 & 0.62 $\pm$0.55 & 61.62 $\pm$22.63 & 73.39 $\pm$20.38 \\
\begin{tabular}[c]{@{}l@{}}DRNN \cite{antczak2018deep}\end{tabular} & 5.85 $\pm$8.93 & 0.44 $\pm$0.30 & 49.91 $\pm$26.92 & 89.48 $\pm$10.28 \\
FCN-DAE \cite{chiang2019noise} & 6.79 $\pm$8.29 & 0.48 $\pm$0.31 & 62.18 $\pm$34.54 & 83.27 $\pm$16.58 \\
Vanilla L & 13.565 $\pm$15.71 & 0.54 $\pm$0.28 & 88.47 $\pm$33.28 & 71.46 $\pm$14.38 \\
Vanilla NL & 6.90 $\pm$9.34 & 0.41 $\pm$0.27 & 63.55 $\pm$38.36 & 85.58 $\pm$11.06 \\
Multibranch LANL & 5.362 $\pm$7.09 & 0.39 $\pm$0.27 & 55.59 $\pm$32.49 & 89.07 $\pm$9.23 \\ \hline
\textbf{Multibranch  LANLD (proposed)} & \textbf{4.29 $\pm$6.35} & \textbf{0.34 $\pm$0.25} & \textbf{45.35 $\pm$29.69} & \textbf{91.46 $\pm$8.61} \\ \hline
\end{tabular}
\end{table}

\begin{figure}[!h]
    \centering
    %\fbox{\rule[-.5cm]{0cm}{4cm} \rule[-.5cm]{4cm}{0cm}}
    \includegraphics[scale=0.39]{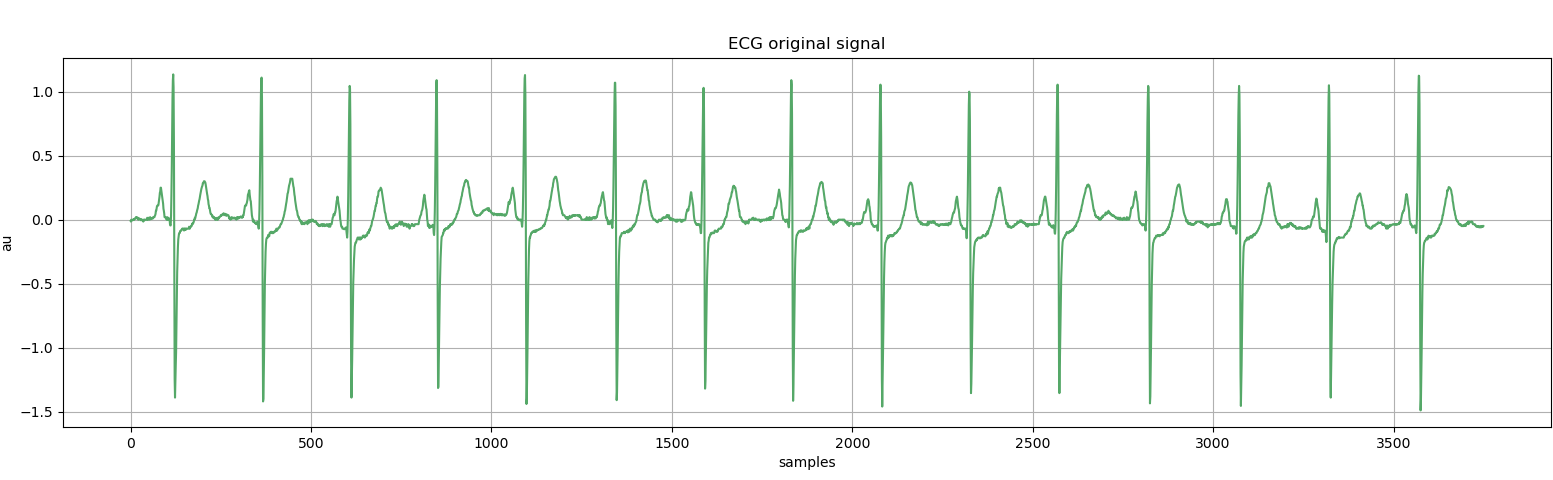}
    \includegraphics[scale=0.39]{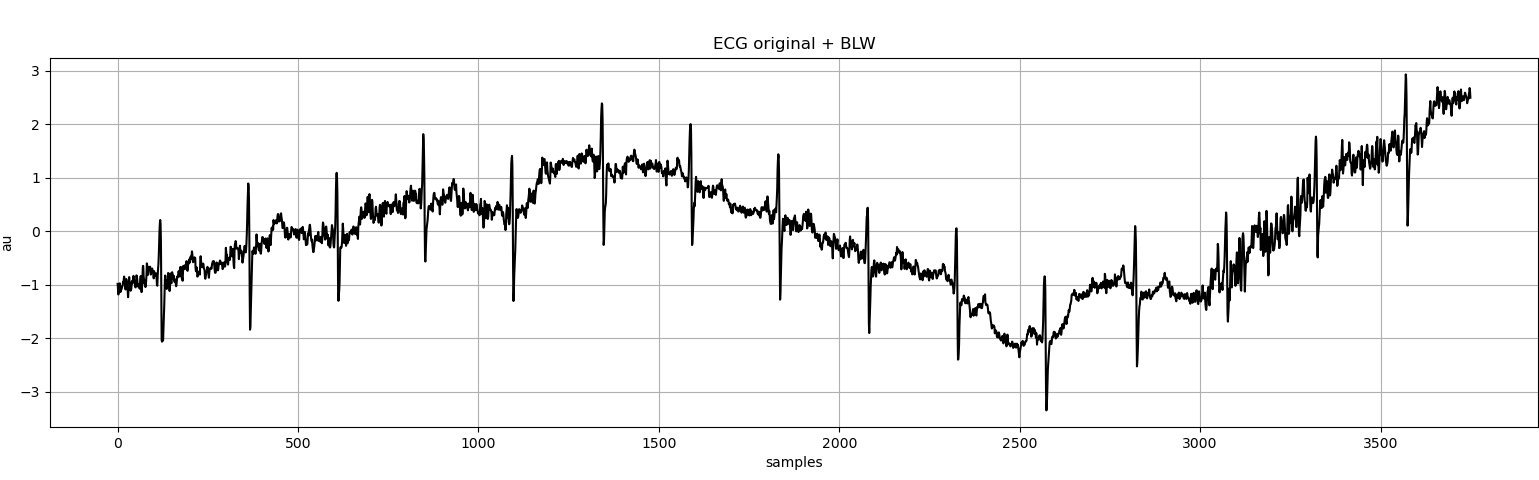}
    \includegraphics[scale=0.39]{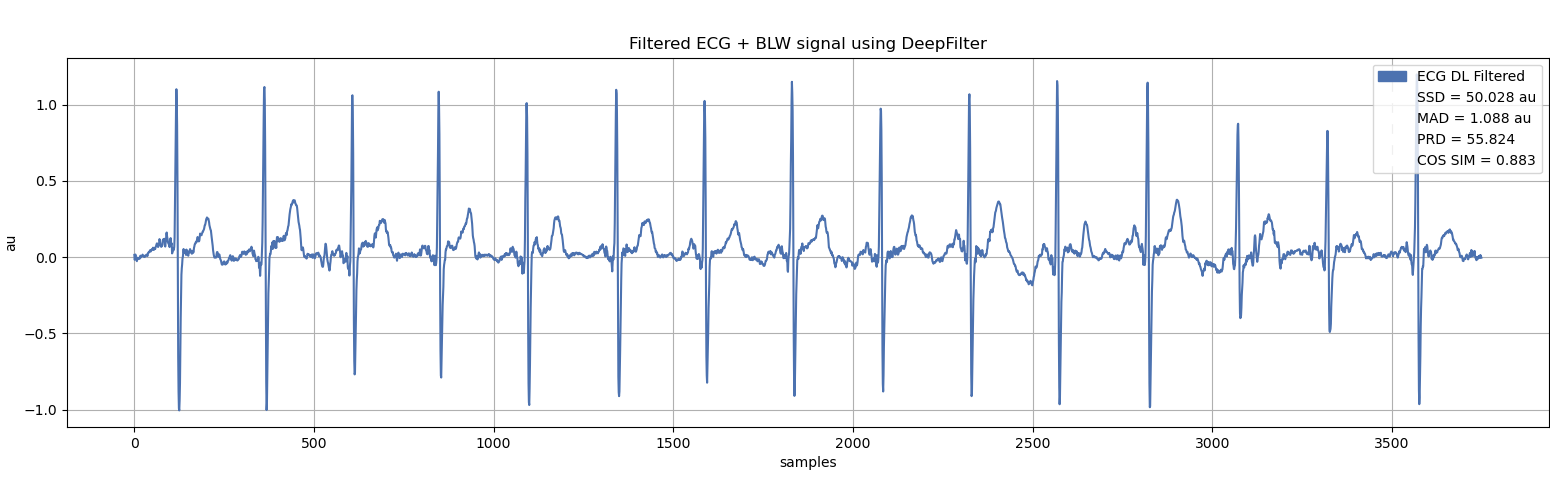}
    \includegraphics[scale=0.39]{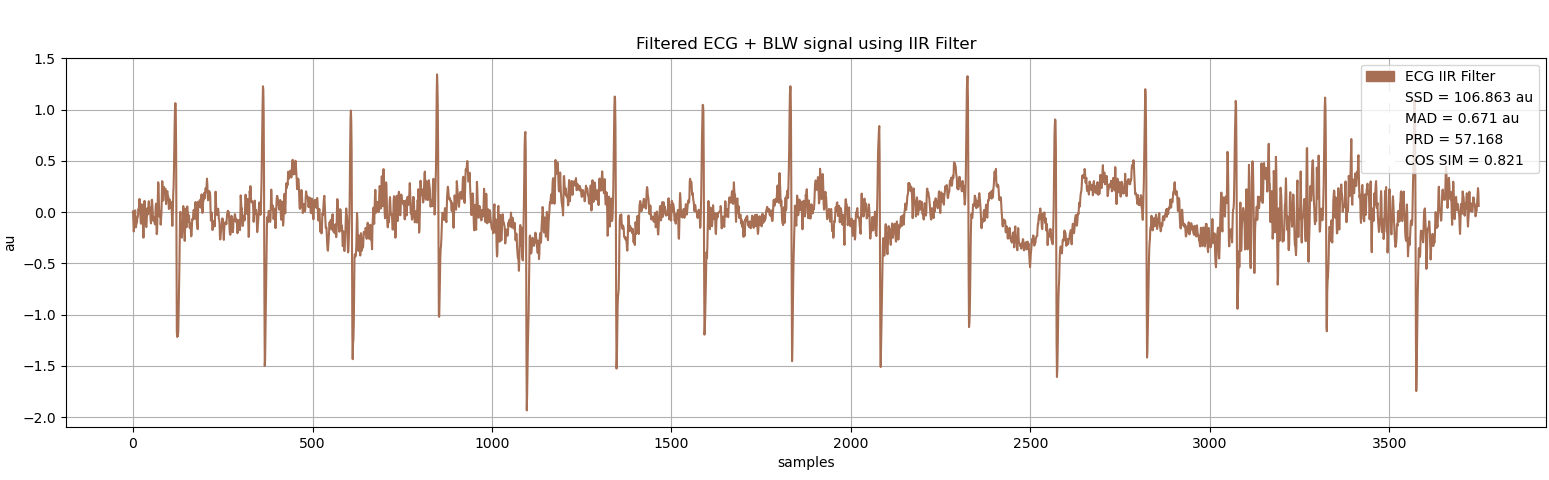}
    \caption{(First) The figure shows a portion of sele0106 ECG signal. (Second) Original ECG + BLW noise from the NSTDB. (Third) The blue line is the ECG filtered using our approach. (Fourth) The brown signal is the ECG recovered using the IIR filter, this image was included for visual comparison purposes. Metric values are also included.}
    \label{fig:qualitative_results}
\end{figure}

Table \ref{tab:test_results} lists the average values and the standard deviation (between parenthesis) yielded by the different methods / models (rows) on the respective metric (columns). When comparing results we can observe that the proposed method obtained better results on all used metrics closely followed by the DRNN \cite{antczak2018deep} model. Despite the DRNN model showcased close results to the proposed model, the difference is  statistically significant (p < 0.01).

When facing a fully linear deep learning model (Vanilla L) against a model that includes non-linear activations (Vanilla NL), we can observe that the model including non-linearities outperforms the fully linear model. Our intuition behind these results is that since models that feature nonlinearities are able to learn more complex functions \cite{goodfellow2016deep} the non-linear model is capable not only of performing a very strong filtering but signal reconstruction as well.

As hypothesized the model featuring dilated convolutions outperforms his counterpart.. As explained previously, dilated convolutions inclearses the filter’s field of view (which is similar to featuring a bigger kernel) allowing the model to better filter low frequency noise.

In Figure \ref{fig:qualitative_results} qualitative and quantitative results for our deep learning approach are shown. On the top an ECG beat from signal sele0106 of the QT Database (in green) . The second signal in black is the same signal + BLW from the NSTDB is shown. The third signal is the ECG signal recovered by our approach is shown in blue line. Finally, on the button the recovered signal by the IIR filter is shown for comparison purposes. Despite the differences between the original signal and the estimated signal by our approach, it can be observed that the signal morphology is highly preserved even when the signal was strongly contaminated by noise.

Another aspect that should be mentioned is the processing speed. Sometimes deep learning approaches tend to be discarded due to high processing time using CPU during deployment.
If it is true that for model training using CPU will take several hours or even days. For final deployment on CPU using the model’s weights obtained during training, the processing time is minimum due to the fact that our model is based on a fully convolutional scheme and also optimizations madel such as dilated convolution. For instance, our model takes ~48 milliseconds to process one ECG beat while the implemented FIR filter takes ~343 milliseconds on the same CPU (Intel(R) Core(TM) i5-7300HQ CPU @ 2.50GHz). Also the proposed approach allows hardware implementation on an application-specific integrated circuit (ASIC) or a field-programmable gate array (FPGA) using a hardware description language (HDL) since convolutional filter modules can be easily parallelizable \cite{ghaffari2020cnn2gate}. This will boost performance allowing the use on real-time applications such as professional ECG monitors.

\section*{Conclusion}

In this work a novel deep learning model for ECG baseline wander removal was presented. The model was compared with other state-of-the-art methods and others classical filtering approaches using similarity metrics. An ablation study to determine the importance of linear and non-linear activations and the use of dilated convolutions was performed. The experimental results showed that the combination of linear and non-linear activation and dilated convolutions on a multi kernel approach proved to be the best solution outperforming state of the art deep learning solutions and classical approaches. 
In order to maximize reproducibility the source code of this work will be freely available in GitHub [\href{https://github.com/fperdigon/DeepFilter}{https://github.com/fperdigon/DeepFilter}]. Also for educational purposes this work can be run on Google Colab where free GPU interactive sessions using jupyter notebooks can be accessed [\href{https://colab.research.google.com/drive/1S1HjkQnrA0EbEDJFr0D-6DAt62RCcP5_?usp=sharing}{Google Colab link}]. Since it is open source future research can use our benchmark to test their own approaches and easily compare with all the models presented in this work (including our implementation of other authors' approaches).

\section*{Supplementary material}
Code hosted in Github:
[\href{https://github.com/fperdigon/DeepFilter}{https://github.com/fperdigon/DeepFilter}]

Run the codes using GPUs at Google Colab:
[\href{https://colab.research.google.com/drive/1S1HjkQnrA0EbEDJFr0D-6DAt62RCcP5_?usp=sharing}{Google Colab link}]

\section*{Credit authorship contribution statement}
\textbf{Francisco Perdigon Romero}: Conceptualization, Methodology, Software, Formal analysis, Investigation, Visualization, Data curation, Writing - review \& editing.\\
\textbf{David Castro Piñol:} Data curation, Writing - review \& editing, Software, Validation.\\
\textbf{Carlos R. Vazquez Seisdedos:} Formal analysis, Writing - review \& editing.

\section*{Declaration of Competing Interest}
The authors declare that they have no known competing financial interests or personal relationships that could have appeared to influence the work reported in this paper.

\bibliographystyle{unsrt}  
\bibliography{references}  

\end{document}